\documentclass[aps,pre,twocolumn,amsmath,amssymb,floatfix,footinbib,superscriptaddress,letterpaper]{revtex4-1}

\usepackage{amsmath,amssymb,mathtools}
\usepackage{stmaryrd}
\usepackage{times}
\usepackage[euler]{textgreek}
\usepackage[bbgreekl]{mathbbol}
\usepackage{graphicx}
\usepackage{bm,color}
\usepackage{natbib,hyperref}
\usepackage[capitalize]{cleveref}

\hypersetup{colorlinks=true,linkcolor=blue,citecolor=blue,urlcolor=blue}

\begin{document}
\title{Entangling light field with mechanical resonator at high temperature}
\author{Qing Lin}
\thanks{These authors contributed equally to this work}
\affiliation{Department of Physics, University of Arkansas, Fayetteville, AR 72701, USA}
\affiliation{Fujian Provincial Key Laboratory of Light Propagation and Transformation, College of Information Science 
and Engineering, Huaqiao University,
Xiamen 361021, China}
\author{Bing He}
\thanks{These authors contributed equally to this work}
\affiliation{Department of Physics, University of Arkansas, Fayetteville, AR 72701, USA}
\author{Liu Yang}
\affiliation{College of Automation, Harbin Engineering University, Heilongjiang 150001, China}
\author{Min Xiao}
\affiliation{Department of Physics, University of Arkansas, Fayetteville, AR 72701, USA}
\affiliation{National Laboratory of Solid State Microstructures and School of Physics, Nanjing University, Nanjing 210093, China}

\begin{abstract}
We present a study on how to realize the widely interested optomechanical entanglement at high temperature. Unlike the majority of the previous experimental and theoretical researches that consider the entanglement of a mechanical resonator with a cavity field created by 
red-detuned continuous-wave or blue-detuned pulsed driving field, we find that applying blue-detuned continuous-wave pump field 
to cavity optomechanical systems can achieve considerable degrees of quantum entanglement, which is generally challenging to obtain at high temperature for the known physical systems. The competition between the induced squeezing-type interaction and the existing decoherence leads to stable entanglement in dynamically unstable regime. There is a much more relaxed condition for the existence of entanglement, as compared with the well-known criterion for neglecting the thermal decoherence on optomechanically coupled systems. 
A simple relation about a boundary in the parameter space, across which the entanglement can exist or not, is found with an analytical expression for the degree of the achieved entanglement at any temperature, which is derived for the systems of highly resolved sideband. 
The studied scenario with blue-detuned continuous-wave driving field can greatly simplify the generation of the widely 
interested optomechanical entanglement of macroscopic quantum states. Our study also provides the answers 
to two fundamentally meaningful open problems: (1) what is the condition for a system to avoid its loss of quantum entanglement under thermal decoherence? (2) is it possible to preserve the entanglement in a thermal environment by increasing the interaction that entangles the subsystems?
\end{abstract}

\maketitle
\section{Introduction}

A majority of realistic quantum systems are open ones coupled to their environment. The influence from the environment, which is generally termed as the ``reservoir", exhibits the overall effect known as decoherence \cite{decoherence,boundary1,boundary2}, leading to the phenomena such as entanglement sudden death (ESD) \cite{y-e1,y-e2}, i.e. the entanglement of a quantum system will be killed by the environment's decoherence after a finite period of time. Especially, the decoherence from thermal environment is significant to quantum systems, 
so that most of quantum information processing systems should operate at ultra-low temperature \cite{ca2,ti1,ti2,ti3,suc1,suc2,suc3}, 
except for a few purely optical ones (see, e.g. \cite{op1,op2,op3}). Possibly preserving quantum entanglement at high temperature is important to both the fundamental researches and potential applications. 

\begin{figure}[b!]
\centering
\includegraphics[width=\linewidth]{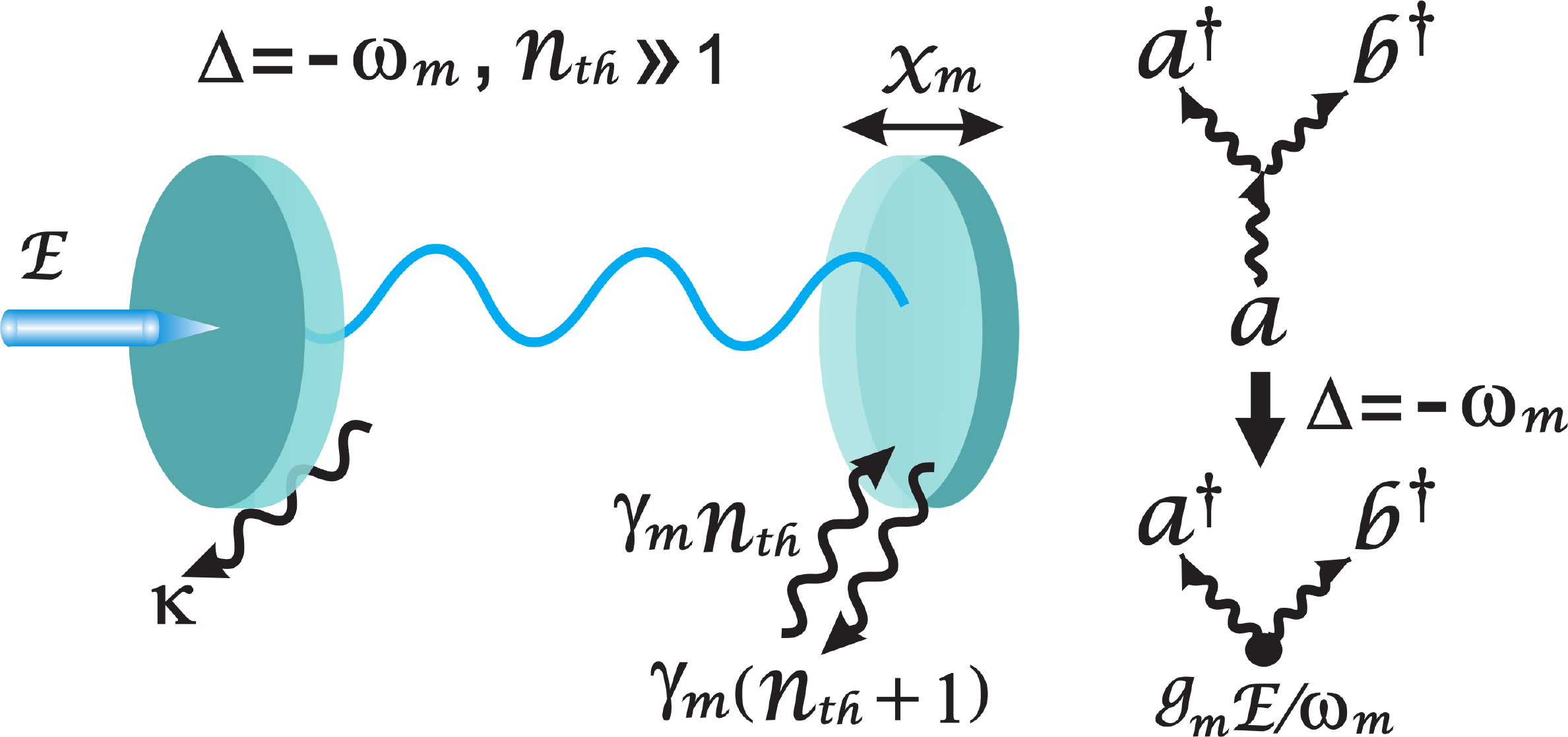}
\caption{Example of optomechanical system. Here the system is driven by a blue-detuned pump field with the intensity $E$ 
and the detuning $\Delta$ at the resonant point $-\omega_m$, so that the nonlinear interaction due to the 
radiation pressure on such a weakly coupled system ($g_m/\omega_m\ll 1$) can be well approximated by a two-mode squeezing action with the intensity $g_mE/\omega_m$. The mechanical resonator is initially in a thermal equilibrium with the environment of the occupation $n_{th}\gg 1$.}
\label{figoms}
\end{figure}
     
Cavity optomechanical system (OMS) is a type of few-body systems that can possibly realize the macroscopic quantum states of a nanomechanical resonator and may find various potential applications \cite{opc1, opc4, opc5}.
A simplified model of the systems is depicted in Fig. 1, while they may have various different realizations \cite{opc5}. 
The entanglement of the nanomechanical resonator with the cavity field has been a widely interested phenomenon since a decade ago (see e.g. \cite{cla1,cla2,cla3,cla4,cla5,cla7,cla8,cla9}). But the generation of such entanglement is a non-trivial task. 
A reported experimental realization of the entanglement \cite{entangle} follows the scenario of pulsed optomechanics \cite{deen,pulsed1,pulsed2,pulse3}. That is to apply a red-detuned pulse first to cool the mechanical resonator, and then entangle the mechanical resonator with another blue-detuned pulse. The entanglement is verified with one more red-detuned pulse which swaps the optomechanical entanglement to the entanglement between two pulsed fields. The whole entanglement generation and verification procedure was performed at a low temperature $T<20$ mK. How to overcome the low temperature restriction and make optomechanical entanglement at higher temperature is significantly meaningful to the relevant experimental researches.

In an OMS, the mechanical resonator with its frequency $\omega_m$ and damping rate $\gamma_m$ undergoes a significant thermal decoherence at high temperature, as the thermal decoherence rate $R(n)=(n+1)n_{th}\gamma_m+n(n_{th}+1)\gamma_m$ \cite{opc5, book} for a Fock state component $|n\rangle$ of the mechanical quantum state grows linearly with the energy level $n$ and 
the thermal occupation $n_{th}=(e^{\hbar\omega_m/(k_B T)}-1)^{-1}$ ($\hbar$ is the Planck's constant and $k_B$ the Boltzmann constant) of the reservoir at the temperature $T$. To neglect the thermal decoherence over one period of mechanical oscillation, the mechanical resonator with the quality factor $Q=\omega_m/\gamma_m$ should meet the relation $Q\omega_m/(2\pi)>k_BT/h$ \cite{opc5}. The correspondingly generalized condition, $Q/n_{th}\gg 1$, should be satisfied for an OMS to be unaffected from the decoherence of its thermal environment (see, e.g. \cite{boundary3,deen,deen1}). Intuitively this condition should also hold for the existence of any quantum property of OMS including the entanglement between cavity field and mechanical resonator. We will however prove below that the condition for the existence of optomechanical entanglement in thermal environment can be actually much more relaxed even possibly to the extent of $Q/n_{th}< 1$, given a suitable coupling between the cavity field and mechanical resonator.

Besides the studies with some specific systems (see, e.g. \cite{pulse3, ex1, ex2, ex3}), the realization of entanglement at higher temperature has been explored with the general systems of coupled harmonic oscillators \cite{h-t,h-t-1,h-t-2}. Those previous studies \cite{h-t,h-t-1,h-t-2} indicates that such entanglement should be generated with time-dependent interaction and out of thermal equilibrium with the environment. For any system, entanglement is created by the mutual interaction between its subsystems. If one simply makes the interaction stronger (the interaction can be time-independent too), will it be possible for the entanglement to survive in an environment of higher temperature? Previously there had been no quantitative results regarding this conjecture. In this work, we will tackle the problem with the example of OMS, showing the possibility to offset the decoherence on entanglement with a proper coupling between cavity field and mechanical resonator. 
The entanglement at very high temperature is found to exist to the dynamically unstable OMSs, which evolve from the initial thermal equilibrium with their environments.

\section{Quantum optomechanical system}
\label{sec:model}

Most previous theoretical descriptions of weakly coupled quantum OMS are based on 
a linearization of the system's nonlinear dynamics by a procedure described as follows (see, e.g. \cite{deen, cla1,cla2,cla3,cla4,cla5, cla7,cla8,cla9}): (1) shifting the cavity field mode $\hat{a}\rightarrow \hat{a}+\alpha$ with respect to its mean-field 
value $\alpha(t)$ and the mechanical mode $\hat{b}\rightarrow \hat{b}+\beta$ with respect to its average displacement $\beta(t)$; (2) neglecting the resulting nonlinear terms treated as the higher-order terms of the fluctuations around the mean values $\alpha(t)$ and $\beta(t)$. In the linearized Hamiltonian obtained through the procedure, the effective coupling intensity of the cavity field with the mechanical resonator will become $g_m|\alpha|=g_m\sqrt{n_p}$, where $g_m$ is the optomechanical coupling intensity at the single-photon level and $n_p$ the average cavity photon number. However, except for the time-independent steady-state solutions, the exact time-dependent mean values $\alpha(t)$, $\beta(t)$ are generally difficult to come by. The previous treatment of pulsed drive field usually adopted certain approximations for these time-dependent averaged values \cite{deen, entangle}. For the entanglement due to continuous-wave (CW) drives, 
most previous theoretical studies of optomechanical entanglement 
(see, e.g. \cite{cla1,cla2,cla3,cla4,cla5,cla7,cla8,cla9}) had to consider the red-detuned ones, because in the blue-detuned regime the steady states exist only for very weak optomechanical coupling \cite{limit}. Blue-detuned CW drive that gives rise to dynamical instability poses a fundamental limit on the applicability of the previous method based on steady states \cite{cla4}. As it has been demonstrated by experiment \cite{entangle}, blue-detuned drive field is the main cause of optomechanical entanglement. The quantum entanglement between cavity field and mechanical resonator can surely exist, even when an OMS is not dynamically stable under blue-detuned CW drive. 

To deal with the situations of blue-detuned CW drives, we adopt a method developed from the approach in \cite{fqa1}, which does not require the mean values $\alpha(t), \beta(t)$ of the system modes in the calculations of the physical quantities. Its improvement over the original approach \cite{fqa1} is to incorporate the fact that the action due to the coupling of the system with its reservoirs should be irreversible and cannot be regarded as a unitary one. Our staring point is from the initial quantum state 
\begin{align}
\rho(0)=\left\vert 0\right\rangle_c\left\langle 0\right\vert \bigotimes \sum_{n=0}^{\infty} \frac{n_{th}^n}{(1+n_{th})^{n+1}}\left\vert n\right\rangle_m\left\langle n\right\vert
\label{ini}
\end{align}
of a concerned OMS, as the product of a cavity vacuum state and a mechanical thermal state before turning on the drive field. Here the mechanical reservoir with the occupation $n_{th}$ can be at arbitrary temperature, as long as the resonator is prepared under the thermal equilibrium with its environment. 

An important feature of an OMS is that it is an open quantum system. The cavity and mechanical thermal reservoir of the OMS can be modeled as the ensembles of oscillators with the continuous distributions of their frequencies (the notation $\hbar=1$ is used from now on):
\begin{eqnarray}
H_R=\int_0^\infty d\omega_1\omega_1\hat{\xi}^\dagger_c\hat{\xi}_c(\omega_1)+\int_0^\infty d\omega_2\omega_2\hat{\xi}^\dagger_m\hat{\xi}_m(\omega_2),
\end{eqnarray}
with $[\hat{\xi}_{c,m}(\omega), \hat{\xi}^\dagger_{c,m}(\omega')]=\delta(\omega-\omega')$.
The coupling of the OMS with the environment modeled by the above Hamiltonian takes the general form
\begin{eqnarray}
H_{SR}&=&i\int d\omega_1 \kappa(\omega_1)(\hat{a}-\hat{a}^\dagger)\{\hat{\xi}^\dagger_c(\omega_1)+\hat{\xi}_c(\omega_1)\}\nonumber\\
&+&i\int d\omega_2 \gamma_m(\omega_2)(\hat{b}-\hat{b}^\dagger)\{\hat{\xi}^\dagger_m(\omega_2)+\hat{\xi}_m(\omega_2)\}.
\end{eqnarray}
The action of the system-reservoir coupling is irreversible, giving rise to the damping of the cavity (mechanical) mode at the rate $\kappa$ ($\gamma_m$). For the smooth coupling between the system and reservoirs, the coupling intensities can be reduced to the constants, i.e. $\kappa(\omega)\rightarrow \sqrt{\kappa/\pi}$, 
$\gamma_m(\omega)\rightarrow \sqrt{\gamma_m/\pi}$ \cite{book}.

The physical processes with the OMS include a drive from an external pump field as well as the coupling between the cavity field and mechanical resonator due to the radiation pressure. In a rotating frame with respect to the external drive frequency $\omega_l$, the Hamiltonians for these processes read \cite{opc5}
\begin{eqnarray}
H_S&=&\Delta \hat{a}^{\dag}\hat{a}+\omega_m \hat{b}^{\dag}\hat{b}
+iE(\hat{a}^{\dag}-\hat{a})
\end{eqnarray}
and
\begin{eqnarray}
H_{OM}=-g_{m}\hat{a}^{\dag}\hat{a}(\hat{b}+\hat{b}^{\dag}),
\end{eqnarray}
where $\Delta=\omega_c-\omega_l$ is the detuning of the drive field's central frequency. 
The drive intensity can be a time-dependent one $E(t)$ to include pulsed fields. 
All physical properties of a quantum OMS are the consequence of the evolution from its initial quantum state Eq. (\ref{ini}) under the action 
$$U(t)={\cal T}\exp\{-i\int_0^t d\tau(H_S+H_{OM}+H_{SR}+H_R)(\tau)\}$$ 
of the total Hamiltonian including that of the stochastic part $H_{SR}$ \cite{book}. The quantum state $\rho_r$ 
of the reservoirs is assumed to be invariant under the action, since the environment is unaffected by the concerned system. 

Next we take an interaction picture with respect to the Hamiltonian $H_S+H_R$. 
The corresponding unitary transformation $U_0(t)=e^{-i(H_S+H_R)t}$ 
(an ordinary exponential due to the commutativity of $H_S+H_R$ at different time) will transform all involved operators 
to
\begin{eqnarray}
&&\hat{a}\rightarrow \hat{A}(t)= U^\dagger_0(t)\hat{a}U_0(t)=e^{-i\Delta t}(\hat{a}+D(t)),\nonumber\\
&& \hat{b} \rightarrow U^\dagger_0(t)\hat{b}U_0(t)=e^{-i\omega_m t}\hat{b},\nonumber\\
&&\hat{\xi}_c(\omega_1)\rightarrow e^{-i\omega_1 t}\hat{\xi}_c(\omega_1),\nonumber\\
&&\hat{\xi}_m(\omega_2)\rightarrow e^{-i\omega_2 t}\hat{\xi}_m(\omega_2),
\label{IO}
\end{eqnarray}
having the displacement
$$D(t)=\int_0^t d\tau e^{i\Delta(t-\tau)}E(\tau)$$ 
for the cavity mode. Correspondingly the rest of the total Hamiltonian becomes 
\begin{align}
H_{eff}(t)=& U^\dagger_0(t)(H_{OM}+H_{SR})U_0(t)\nonumber\\
=&-g_m\left[D(t)\hat{a}^{\dag}+D^*(t)\hat{a}+|D(t)|^2\right]\nonumber\\
&\times(e^{-i\omega_mt}\hat{b}+e^{i\omega_mt}\hat{b}^{\dag}) \nonumber\\
&\underbrace{-g_m \hat{a}^\dagger\hat{a}(e^{-i\omega_mt}\hat{b}+e^{i\omega_mt}\hat{b}^{\dag})}_{H_{N}(t)}\nonumber\\
&+i\sqrt{2\kappa}\{e^{-i\Delta t}\hat{A}^{\dag}(t) \hat{\xi}_c(t)-e^{i\Delta t}\hat{A}(t)\hat{\xi}^{\dag}_c(t)\} \nonumber\\
&+i\sqrt{2\gamma_m}(\hat{b}^{\dag}\hat{\xi}_m(t)-\hat{b}\hat{\xi}^{\dag}_m(t)), 
\label{effective}
\end{align}
where 
\begin{eqnarray}
\hat{\xi}_{c}(t)&=&\frac{1}{\sqrt{2\pi}}\int d\omega \hat{\xi}_{c}(\omega)e^{-i(\omega-\Delta)t},\nonumber\\
\hat{\xi}_{m}(t)&=&\frac{1}{\sqrt{2\pi}}\int d\omega \hat{\xi}_{m}(\omega)e^{-i(\omega-\omega_{m})t}.
\label{n-operator}
\end{eqnarray}
In the above Eq. (\ref{effective}) we have applied the rotating-wave approximation to neglect the terms containing the fast oscillating factors in the system-reservoir coupling part. For the weak system-reservoir interactions leading to the relatively slow coupling processes, the associated quantum noise operators can also be treated as the white-noise operators satisfying $\langle \hat{\xi}^{\dag}_l (t) \hat{\xi}_l (t') \rangle_R=n_l\delta(t-t')$ and $\langle \hat{\xi}_l (t) \hat{\xi}_l^{\dag} (t') \rangle_R=(n_l+1)\delta(t-t')$, for $l=c, m$ with the respective occupation $n_l$ \cite{book}. In what follows we will consider a vanishing thermal occupation for the cavity reservoir, 
which is appropriate to pump drive fields with optical frequencies. 

The essential point in our approach is a proper use of the effective Hamiltonian, Eq. (\ref{effective}), for finding the expectation values of the evolved system operators (including those as the observables in experiments).
The expectation value of an arbitrary operator $\hat{O}(t)=U^\dagger(t)\hat{O}U(t)$ that evolves with time can be written as 
\begin{widetext} 
\begin{eqnarray}
\langle \hat{O}(t)\rangle&=&\mbox{Tr}_S\{\hat{O}\rho(t)\}=\mbox{Tr}_S\{\hat{O}\mbox{Tr}_R(U(t)\rho(0)\rho_r U^\dagger(t))\}\nonumber\\
&=&\mbox{Tr}_{S,R}\{\hat{O}U_0(t){\cal T}e^{-i\int_0^t d\tau H_{eff}(\tau)}\rho(0)\rho_r {\cal T}e^{i\int_0^t d\tau H_{eff}(\tau)}U^\dagger_0(t)\}\nonumber\\
&=& \mbox{Tr}_{S,R}\{{\cal T}e^{i\int_0^t d\tau U_N(t,\tau)(H_{eff}-H_N)(\tau)U^\dagger_N(t,\tau)}U^\dagger_0(t)\hat{O}U_0(t){\cal T}e^{-i\int_0^t d\tau U_N(t,\tau)(H_{eff}-H_N)(\tau)U^\dagger_N(t,\tau)}\nonumber\\
&\times & U_N(t,0)\rho(0)\rho_r U_N^\dagger(t,0)\}\nonumber\\
&\approx& \mbox{Tr}_{S,R}\{{\cal T}e^{i\int_0^t d\tau (H_{eff}-H_N)(\tau)}U^\dagger_0(t)\hat{O}U_0(t){\cal T}e^{-i\int_0^t d\tau (H_{eff}-H_N)(\tau)}\rho(0)\rho_r\},
\label{expect}
\end{eqnarray}
\end{widetext}
in which the trace is taken over both the system part $S$ and the reservoir part $R$. As emphasized in \cite{book}, the action 
$U(t)$ in the above is only a formally unitary one, since it also involves the action of the system-reservoir couplings 
that cause the damping of the system modes as the non-unitary processes. The interaction picture with $U_0(t)$ as mentioned above is equivalent to the factorization of the operator $U(t)$ on the second line of Eq. (\ref{expect}). To determine the effect of the nonlinear term $H_N(t)$ in Eq. (\ref{effective}), we continue to factorize its action $U_N(t,0)={\cal T}\exp\{-i\int_{0}^t dt'H_N(t')\}$ out of the evolution operator ${\cal T}e^{-i\int_0^t d\tau H_{eff}(\tau)}$ as 
on the third and fourth line of Eq. (\ref{expect}). The operation $U_N(t,0)$ factorized out in Eq. (\ref{expect}) keeps the initial quantum state $\rho(0)\rho_r$ invariant since the cavity is initially in a vacuum state $|0\rangle_c$. 
In both of the factorization procedures, we only perform the truly unitary operations to modify the system mode operators in the remaining Hamiltonians, to avoid the actually non-unitary action by the system-reservoir coupling $H_{SR}$ used in \cite{fqa1}. The second factorization modifies the operators in the remaining Hamiltonian $H_{eff}(\tau)-H_N(\tau)$ by a unitary operation $U_N(t,\tau)={\cal T}\exp\{-i\int_{\tau}^t dt'H_N(t')\}$ \cite{fqa1}:
\begin{eqnarray}
&&\hat{A}_I(\tau)=U_N(t,\tau)\hat{a}U^\dagger_N(t,\tau)\nonumber\\
&=&\exp\{i\int_\tau^t dt'\frac{g^2}{\omega_m}(1-e^{i\omega_m(t'-\tau)})(\hat{a}^\dagger\hat{a}+1) \}\nonumber\\
&\times & \exp\{-i\frac{g_m}{\omega_m}(e^{i\omega_m t}-e^{i\omega_m\tau})\hat{B}_I^\dagger(\tau)\}\nonumber\\
&\times & \exp\{i\frac{g_m}{\omega_m}(e^{-i\omega_m t}-e^{-i\omega_m\tau})\hat{b}(\tau)\}\hat{a};\nonumber\\
&&\hat{B}_I(\tau)=U_N(t,\tau)\hat{b}U^\dagger_N(t,\tau)\nonumber\\
&=&\hat{b}-\frac{g_m}{\omega_m}(e^{i\omega_m t}-e^{i\omega_m\tau})\hat{a}^\dagger\hat{a}.
\label{correct}
\end{eqnarray}
These modified operators differ from the original ones, $\hat{a}$ and $\hat{b}$, 
by the terms in the orders of $g_m/\omega_m$ from the expansions of the above equations, and such corrections can be neglected for a weakly coupled OMS satisfying $g_m/\omega_m\ll 1$ [especially in the highly resolved sideband regime $\omega_m/\kappa\gg 1$, considering the oscillation factors in Eq. (\ref{correct})]. Throughout our derivations, the approximation sign on the fifth line of Eq. (\ref{expect}) due to this practice is the only step that is not exact, in addition to the commonly used rotating-wave approximation and others to derive the system-reservoir coupling Hamiltonian. 

Finding the value of $\langle \hat{O}(t)\rangle$ can be therefore reduced to determining the evolved operator $\hat{O}(t)$ due to the successive actions $U_0(t)$ and ${\cal T}e^{-i\int_0^t d\tau (H_{eff}-H_N)(\tau)}$, and then taking its expectation value with respect to the fixed initial quantum state $\rho(0)\rho_r$. The evolved operator $\hat{O}(t)$ may contain the quantum noise operators, whose averages over the total reservoir state $\rho_r$ should be found by the correlation relations of the noise operators in Eq. (\ref{n-operator}). The first action $U_0(t)$ evolves the system operators exactly as in Eq. (\ref{IO}).
The evolutions of the operators $\hat{O}=\hat{a}$ and $\hat{b}$ under the second action  
${\cal T}e^{-i\int_0^t d\tau (H_{eff}-H_N)(\tau)}$ of a quadratic Hamiltonian are determined by the corresponding dynamical equations that are in the exact forms as well. Making use of the proper Ito's rules for the quantum noise operators in the system-reservoir coupling part \cite{book}, one will obtain the dynamical equations 
as follows:
\begin{align}
\dot{\hat{a}}=&-\kappa \hat{a}+i g_m D(t) (e^{-i\omega_mt}\hat{b}+e^{i\omega_mt}\hat{b}^{\dag})-\kappa D(t) \nonumber \\
&+\sqrt{2\kappa}\hat{\xi}_c (t),\nonumber \\
\dot{\hat{b}}=&-\gamma_m \hat{b}+i g_m e^{i\omega_m t}\left[D(t)\hat{a}^{\dag}+D^*(t)\hat{a}\right] \nonumber\\
&+i g_m e^{i\omega_m t}|D(t)|^2+\sqrt{2\gamma_m}\hat{\xi}_m(t).
\label{eq:dm}
\end{align}
These linear differential equations including the coherent and noise drive terms can be numerically solved. As seen from the equations, the effect of the coupling terms of squeezing type, which are proportional to $\hat{a}^\dagger$ or $\hat{b}^\dagger$ on their right sides and contribute to the optomechanical entanglement primarily, can be significantly enhanced by a blue-detuned CW drive set at $\Delta=-\omega_m$. The dominant squeezing effect in this situation manifests more explicitly by the corresponding equations
\begin{align}
\dot{\hat{a}}=&-\kappa \hat{a}+ J\hat{b}^{\dag}+i\kappa g_m^{-1}J +\sqrt{2\kappa}\hat{\xi}_c (t),\nonumber \\
\dot{\hat{b}}=&-\gamma_m \hat{b}+J\hat{a}^{\dag} -ig_m^{-1}J^2+\sqrt{2\gamma_m}\hat{\xi}_m(t)
\label{eq:sdm}
\end{align}
in the limit $\omega_m/\kappa\rightarrow \infty$, to have the constant squeezing coupling terms proportional 
to $J=g_m E/\omega_m$, since the effects of the other coupling terms carrying oscillating factors are completely averaged out in this limit.

\begin{figure}[b!]
\centering
\includegraphics[width=\linewidth]{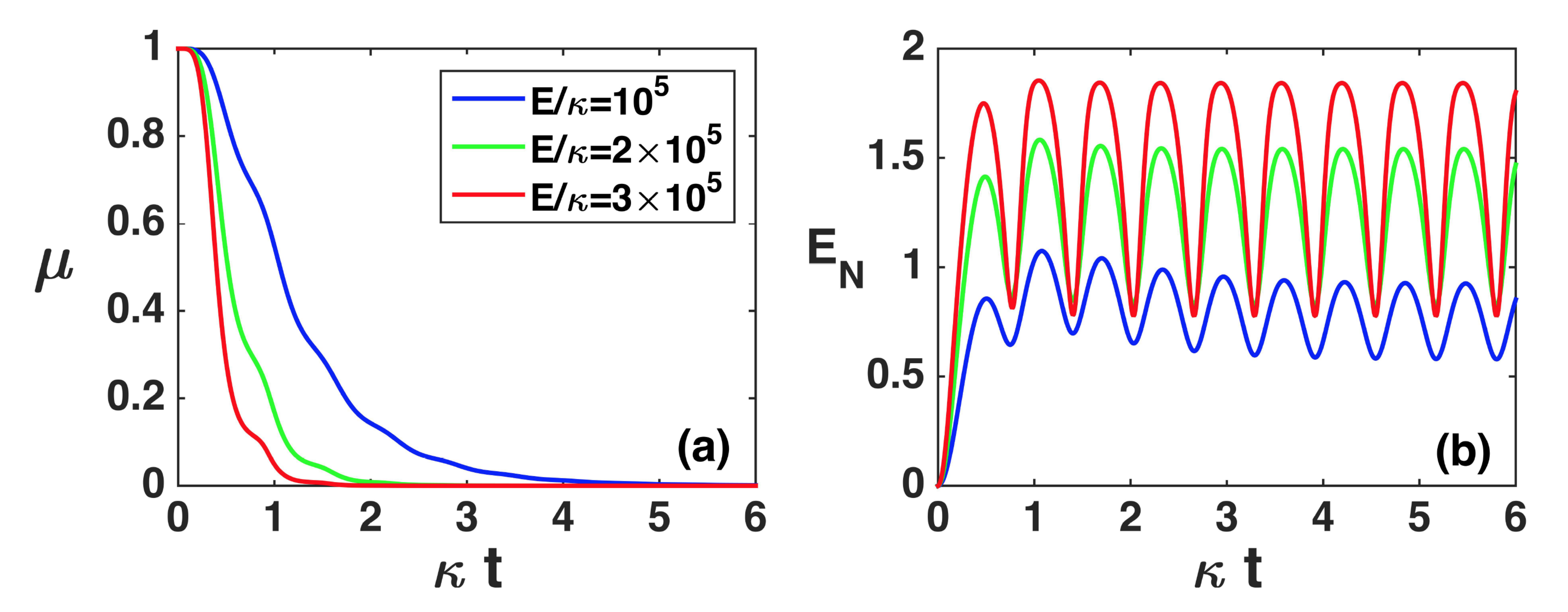}
\caption{(a) Purity evolutions for the quantum states of an OMS at zero temperature. (b) Evolutions of the corresponding entanglement values. Here the OMS is with the system parameters $g_m/\kappa=10^{-4},\omega_m/\kappa=10,Q=10^6$, and the drive's detuning is $\Delta=-\omega_m$. }
\label{fig1}
\end{figure}

\begin{figure*}[t!]
\centering
\includegraphics[width=\linewidth]{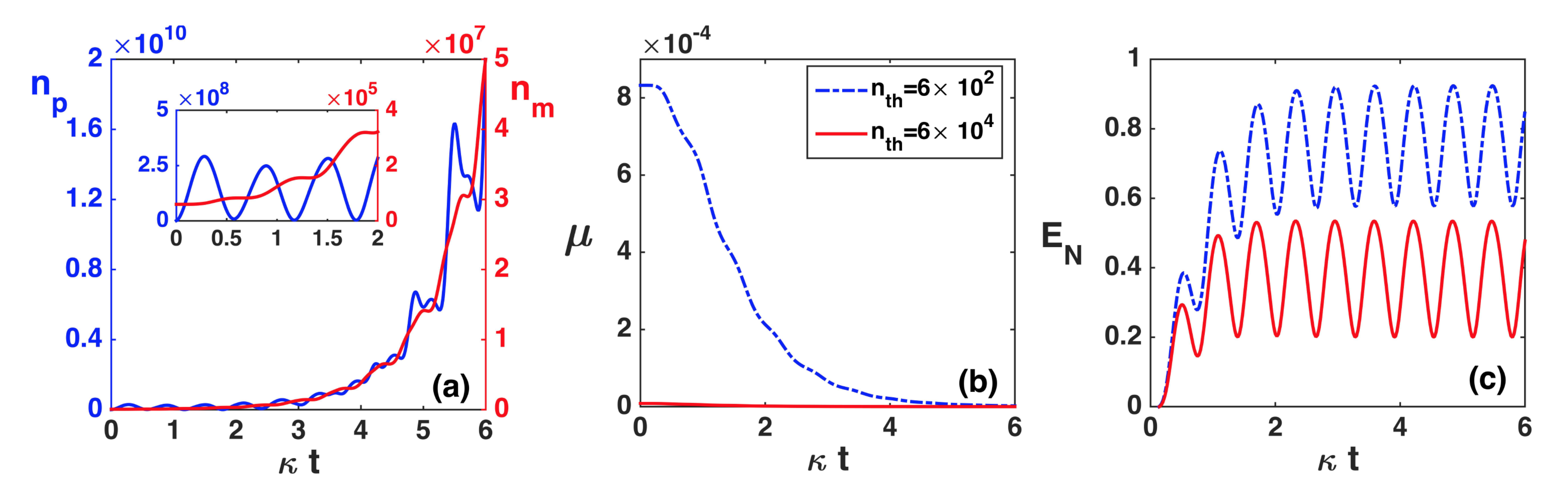}
\caption{Real-time evolutions of cavity photon number, thermal phonon number, purity and entanglement in thermal environment. In (a) we consider the room temperature $T=300$ K for a mechanical resonator with the frequency $\omega_m/2\pi=100$ MHz, and the system is driven by a blue-detuned ($\Delta/\omega_m=-1$) CW pump field with the intensity $E/\kappa=10^5$. The inset in (a) shows that the thermal phonon number evolves from the corresponding initial thermal 
occupation number $n_{th}=6\times 10^4$, while the cavity photon number grows from zero after turning on the drive field. In (b) and (c), two initial temperatures corresponding to the indicated different thermal occupations are used for the evolutions the purity and entanglement. 
The parameters for the system are taken as $g_m/\kappa=10^{-4}$, $\omega_m/\kappa=10$, and $Q=10^6$.}
\label{fig3}
\end{figure*}

Though it is impossible to have dynamical stability for a squeezing dominant situation, the quantum states of an OMS evolved from the initial state $\rho(0)$ in Eq. (\ref{ini}) will still be Gaussian states, if the evolution process can be approximated with the successive actions of $U_0(t)$ and ${\cal T}e^{-i\int_0^t d\tau (H_{eff}-H_N)(\tau)}$ as 
in Eq. (\ref{expect}).  
Such Gaussian states can be depicted by the $4\times 4$ correlation matrix (CM)
\begin{eqnarray}
\hat{V}= \left(
\begin{array}
[c]{cc}%
 \hat{A} & \hat{C} \\
 \hat{C}^T &  \hat{B}
\end{array}
\right),
\label{CM}
\end{eqnarray}
with its elements defined as 
$$V_{ij}(t)=\frac{1}{2}\langle \delta \hat{u}_i(t)\delta \hat{u}_j(t)+\delta\hat{u}_j(t)\delta\hat{u}_i(t)\rangle.$$
The fluctuation $\delta \hat{u}_i(t)=\hat{u}_i(t)-\langle \hat{u}_i(t)\rangle$ considered here 
is around the time-dependent 
expectation value $\langle \hat{u}_i(t)\rangle$ 
for the elements of the vector $\hat{\vec{u}}(t)=(\hat{x}_c(t),\hat{p}_c(t),\hat{x}_m(t),\hat{p}_m(t))^T$, where $\hat{x}_l=(\hat{c}+\hat{c}^{\dag})/\sqrt{2}$, $\hat{p}_l=-i(\hat{c}-\hat{c}^{\dag})/\sqrt{2}$ for $l=c,m$ and $c=a,b$.
As a contrast, in the majority of the previous studies confined to the regimes of 
red-detuned CW drive, the expectation values $\langle \hat{u}_i(t)\rangle$ should be 
time-independent steady ones.
Our concerned time-dependent CM elements for the quantum states of any weakly coupled OMS can be found from Eqs. (\ref{expect}) and (\ref{eq:dm}), directly giving the Wigner functions of its evolving Gaussian states.  
A measure for the corresponding entanglement between the cavity and mechanical modes is the
logarithmic negativity $E_\mathcal{N}=\max [0, -\ln (2 \eta^-)]$ \cite{n1,n2,n3,n4}, where 
\begin{eqnarray}
\eta^-=\sqrt{\Sigma-\sqrt{\Sigma^2-4\det\hat{V}}}/\sqrt{2} 
\label{nega}
\end{eqnarray}
and 
$\Sigma=\det\hat{A}+\det\hat{B}-2\det\hat{C}$. So far we have developed the theoretical tools that can deal with the optomechanical entanglement due to blue-detuned CW drives.

\section{Entanglement of dynamically unstable systems}
\label{sec:dm}

In this section we will present the detailed results of the evolved optomechanical entanglement due to a CW pump laser 
set at the squeezing resonant point $\Delta=-\omega_m$. The other properties of the OMSs never approaching a dynamical stability in the regime will be discussed too. In this situation the effect of the noise drive terms in Eq. (\ref{eq:dm}) will be intensified, so one expects significant decoherence from the cavity and mechanical reservoir, which could eliminate the quantum features of the systems. Meanwhile a significantly enhanced two-mode squeezing effect at $\Delta=-\omega_m$ is surely beneficial to entangling the cavity field and mechanical resonator. The realized entanglement is obviously the result of the competition between these two factors.

We first look at a setup at the idealized situation of $T=0$. Given a high mechanical quality factor the decoherence to this setup predominantly comes from its cavity reservoir, which manifests as the terms proportional to $\sqrt{2\kappa}$ in Eq. (\ref{effective}). Now the initial quantum state in Eq. (\ref{ini}) is reduced to a pure state $|0\rangle_c|0\rangle_m$. A straightforward figure-of-merit for the two-mode system's resistance to the decoherence is how its evolved state 
$\rho(t)$'s purity \cite{p1}
\begin{eqnarray}
\mu(\rho(t))=\mbox{Tr}\rho^2(t)=\frac{1}{4\sqrt{\mbox{Det}\hat{V}(t)}},
\label{purity}
\end{eqnarray}
where $\hat{V}$ is the CM defined in Eq. (\ref{CM}), would survive in the end. Its time evolution is illustrated with three different drive intensities in Fig. \ref{fig1}(a). The initial unit purity will nonetheless tend to zero with time, indicating the complete decoherence even at zero temperature. The corresponding entanglement values for the different drive intensities, however, will all become stably oscillating ones as shown in Fig. \ref{fig1}(b). This example demonstrates the existence of optomechanical entanglement in spite of the dominant decoherence effect that is constantly lowering the purity.   
 
The evolution tendencies of the entanglement and purity can be preserved at non-zero temperature (such as the room temperature) as shown in Fig. 3. For an OMS at the room temperature we plot how its associate thermal phonon number 
\begin{eqnarray}
n_m(t)&=&\langle \hat{b}^{\dag}(t) \hat{b}(t) \rangle -\langle \hat{b}^{\dag}(t) \rangle \langle \hat{b}(t) \rangle\nonumber\\
&=&\langle \delta\hat{b}^{\dag}(t) \delta\hat{b}(t) \rangle,
\end{eqnarray}
where $\delta \hat{b}(t)=\hat{b}(t)-\langle\hat{b}(t)\rangle$, and cavity photon number $n_p(t)=\langle \hat{a}^{\dag}(t) \hat{a}(t) \rangle$ evolve with time. 
The former is purely from the contributions of the noise drive terms and the expectation value of the evolved system mode operators [taken with respect to the initial state (\ref{ini})], since the contribution from the coherent drive terms is subtracted out in the above equation. Its constant increase means the heating of the OMS in such blue-detuned regime, especially under the intensified thermal decoherence from the mechanical reservoir. 
Such growing magnitude of the fluctuation $\delta \hat{b}(t)$, which increases from $\langle \delta\hat{b}^{\dag}\delta \hat{b}(0) \rangle=n_{th}$ to even higher values, 
clearly indicates that the previously adopted linearization by its expansion 
around the mean value $\beta(t)=\langle\hat{b}(t)\rangle$ is not workable in the currently concerned situations. 
In its initial thermal state, the mechanical resonator's fluctuation magnitude $\langle \delta\hat{b}^{\dag}\delta \hat{b}\rangle$ 
has been far above the corresponding mean value $|\beta(0)|^2=0$ already. Meanwhile, the cavity photon number quickly grows with the two-mode squeezing effect, implying that the effective coupling between the cavity field and mechanical resonator keeps being enhanced with time. The evolving entanglement between the cavity field and mechanical resonator will, however, stabilize after a period of time, 
due to the balance of the total decoherence and optomechanical coupling. 
The significant decoherence that exists in the processes manifests by the vanishing purities in Fig. 3(b), though 
their initial values have been rather low in the thermal environment. Under the dominant squeezing effect, the determinant 
of the CM in Eq. (\ref{purity}) will diverge, inevitably eliminating the remnant purity of any initially prepared state in the end. On the other hand, the entanglement magnitude is determined by the factor $\Sigma(t)-\sqrt{\Sigma^2(t)-4\det\hat{V}(t)}$ in Eq. (\ref{nega}). With the progress of an evolution, the subtraction of the two diverging terms in the factor will converge to a stably oscillating function, resulting in the illustrated entanglement.

It is possible to obtain the optomechanical entanglement in the blue-detuned regime with the currently available experimental setups. In Fig. 4 we present the evolved entanglement under two different drives acting on the optomechanical cooling setup in \cite{cool5}, but the setup is operated at the room temperature. The values of the realized stable entanglement are rather high. The averaged entanglement value is larger under the stronger one of the two used drive intensities, since it brings about a higher radiation pressure on the mechanical resonator. 

\begin{figure}[h!]
\centering
\includegraphics[width=\linewidth]{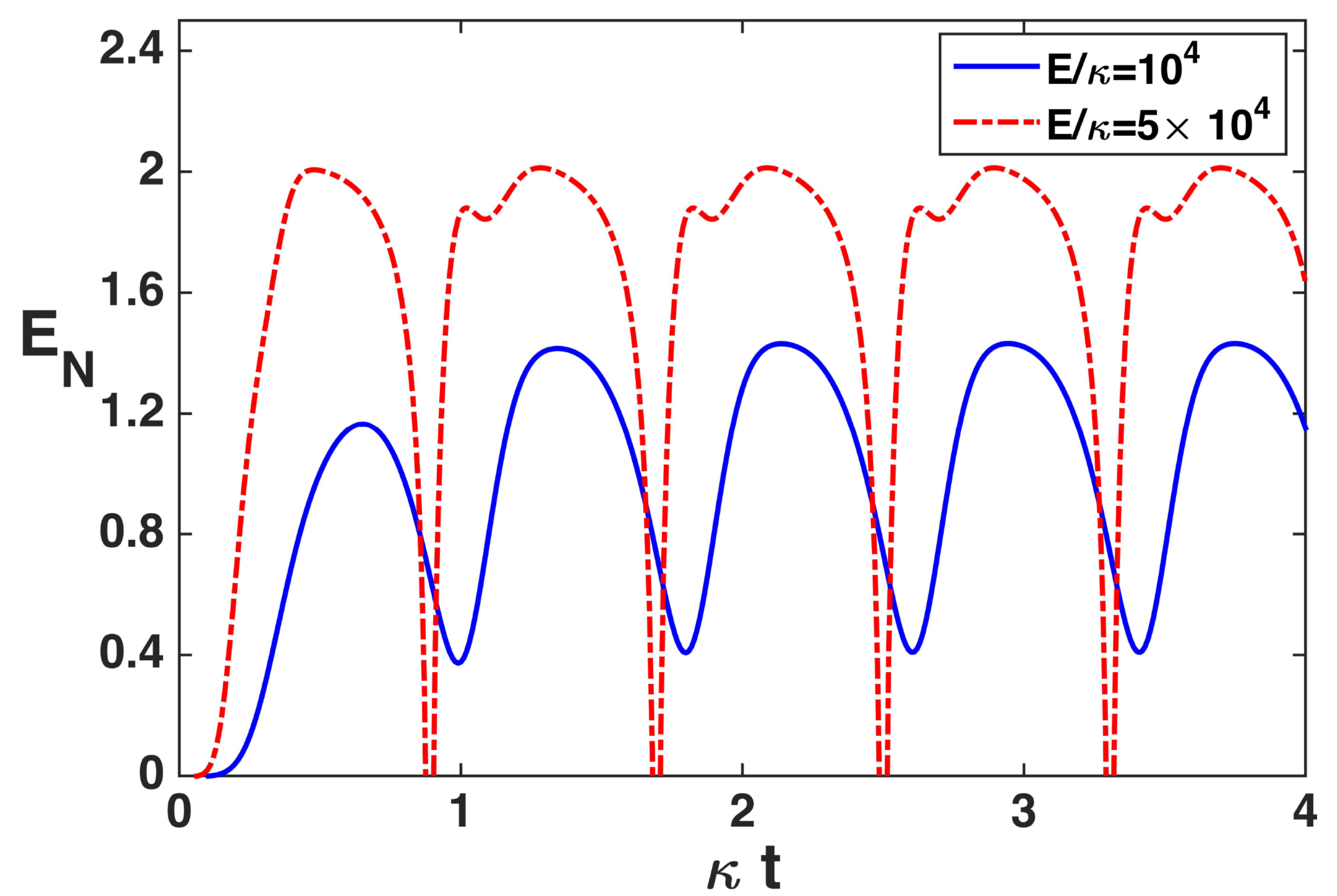}
\caption{Entanglement that can be achieved with the experimental setup in Ref. \cite{cool5}, 
which has the parameters $\omega_m/2 \pi =3\times 10^9$ Hz, $\kappa/2\pi=5\times 10^8$ Hz, $g_m/2\pi=9\times 10^5$ Hz, $Q=10^5$. Here we consider its performance at the room temperature $T=300$ K corresponding to $n_{th}=1.6\times 10^3$.}
\label{figs2}
\end{figure}

\begin{figure}[t!]
\centering
\includegraphics[width=\linewidth]{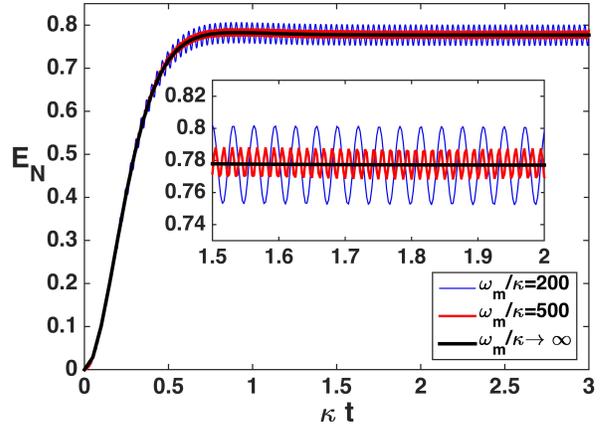}
\caption{Evolved entanglement for the setups with $\omega_m/\kappa \gg 1$. The system parameters are $g_m/\kappa=10^{-4},Q=10^6$. The drive intensity and the thermal reservoir occupation are set to satisfy the relations $J/\kappa=(g_m/\omega_m)(E/\kappa)=2.5$ and $(\gamma_m/\kappa) n_{th}=1$. } 
\label{figs1}
\end{figure}

An interesting phenomenon with the higher drive intensity in Fig. 4 is the significant oscillation of the entanglement value, which leads to the periodic ESD \cite{y-e1,y-e2} and entanglement revival.  
The cause of such oscillation is relevant to the oscillating coupling terms in Eq. (\ref{eq:dm}) in addition to the drive intensity. 
One can see the point by improving on the sideband resolution $\omega_m/\kappa$ for an OMS. In the limit $\omega_m/\kappa\rightarrow \infty$, the oscillating terms in the dynamical equations will take no effect, to reduce the dynamical equations to Eq. (\ref{eq:sdm}). 
The stable entanglement value in this limit takes the analytical form
\begin{align}
E_{\mathcal{N}}= -\ln \left[\frac{J^2(\kappa+2\gamma_m n_{th})-(\kappa\sqrt{4J^2+\kappa^2}-\kappa^2)\gamma_m n_{th}}{J^2\sqrt{4J^2+\kappa^2}}\right] 
\label{analy}
\end{align}
under the realizable condition $\kappa\gg \gamma_m$. As shown in Fig. 5, the oscillating entanglement values numerically calculated according to Eq. (\ref{eq:dm}) will asymptotically tend to this steady result. This completely steady entanglement obtained from Eq. (\ref{eq:sdm}) 
only with the constant coupling terms also distinguishes our conclusions from the previous proposals \cite{h-t, h-t-1,h-t-2} 
that high-temperature entanglement between two harmonic oscillators should be created with time-dependent interaction. 

\section{how entanglement becomes stabilized}
\label{sec:noise}

From the above discussions one sees that optomechanical entanglement under blue-detuned CW drive can exist at sufficiently high temperature. The technically achievable OMSs with their mechanical quality factor $Q=10^5-10^6$ \cite{opc5} provide a specific example that the decoherence effect on the quantum entanglement, especially the thermal one at high temperature, can be overcome by a proper mutual interaction to entangle the sub-systems. We also see from the illustrated examples that the decoherence effect on entanglement is totally different from the corresponding effect on the coherence of a system as the loss of purity. A common feature of the obtained entanglement is its stable oscillation in the concerned dynamically unstable regime. The causes for such uniqueness of the entanglement should be clarified.

\begin{figure}[t!]
\centering
\includegraphics[width=\linewidth]{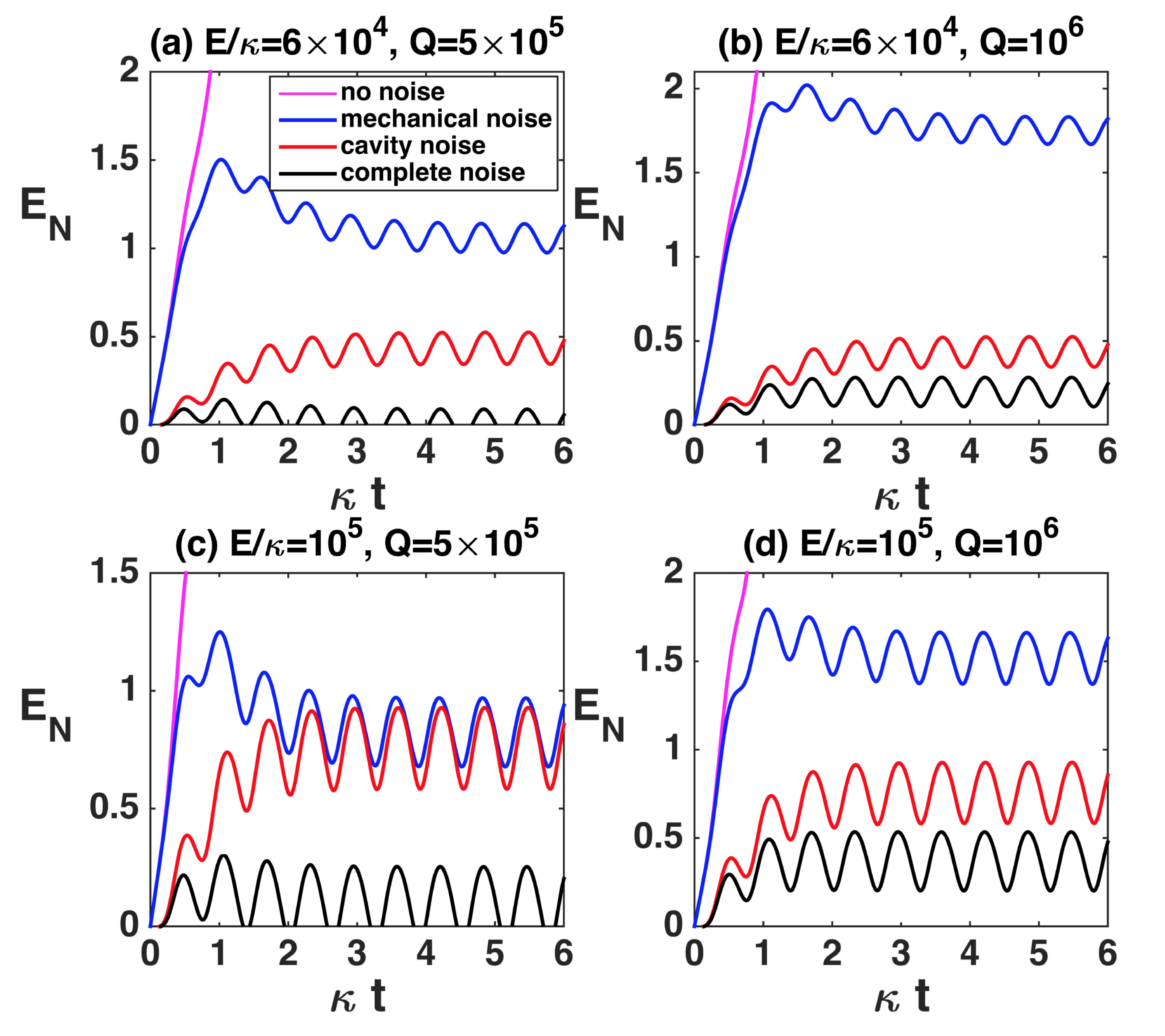}
\caption{Comparison of the entanglement values with and without quantum noise effects. Here we calculate the logarithmic negativity for the entanglement evolved with only the cavity noise, with only the mechanical noise and under both the cavity and mechanical noises, respectively. The other system parameters, which are not indicated for the plots, are the same as those in Fig. 3.}
\label{}
\end{figure}
\vspace{-0cm}

\begin{figure}[b!]
\centering
\includegraphics[width=\linewidth]{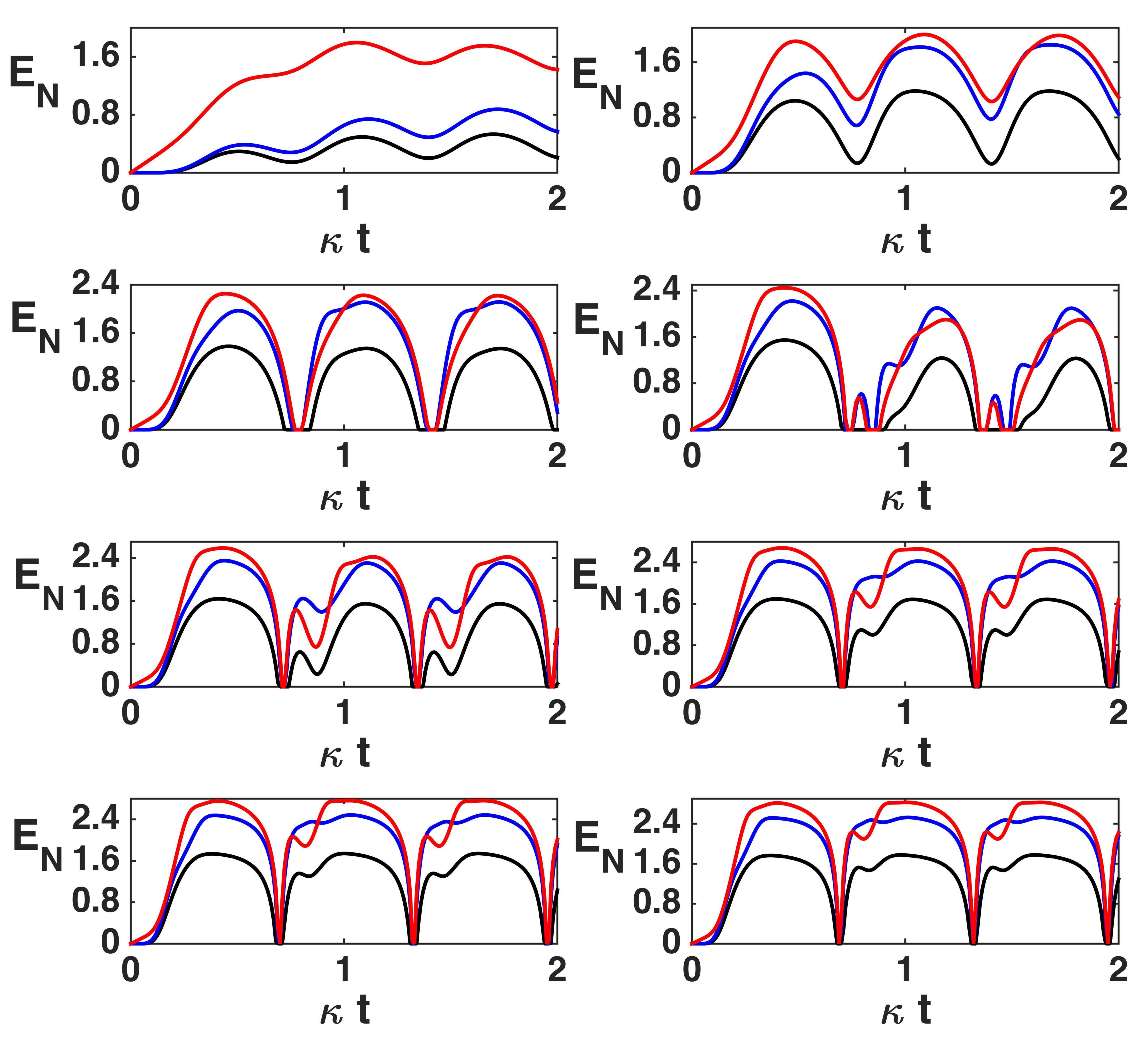}
\caption{Evolved entanglement with a series of gradually increased drive intensity. Following the order from the upper left frame to the lower right frame, the drive intensity $E/\kappa$ grows from $10^5$ to $2.4\times 10^6$ with a gap of $2\times 10^5$ for each step. The system parameters are the same as those in Fig. 6, with the mechanical quality factor fixed as $Q=10^6$. The colors of the curves are in one-to-one correspondence with those in Fig. 6. }
\label{figs3}
\end{figure}

We identify what determine the entanglement dynamics from the dynamical equations, Eq. (\ref{eq:dm}). 
In those equations the decoherence from the cavity and mechanical reservoir manifest as the quantum noise drive terms proportional to $\sqrt{2\kappa}$ and $\sqrt{2\gamma_m}$, respectively. If removing these terms, the associated entanglement 
would evolve according to a dynamics without the decoherence from the environment. Under the enhanced two-mode squeezing due to a blue-detuned CW drive field, the optomechanical entanglement will grow monotonously with time under such assumed dynamics; see the pink curves in Fig. 6. To see the effects of the different types of quantum noise, one can separately add the noise drive terms back to the equations. Only with the mechanical noise term $\sqrt{2\gamma_m}\hat{\xi}_m$ in Eq. (\ref{eq:dm}), the evolved entanglement will become stabilized oscillation. A larger quality factor $Q$ is found to suppress the thermal noise effect further, by comparing the blue curves in Figs. 6(a) and 6(c) with the corresponding ones in Figs. 6(b) and 6(d). The effect of the cavity noise, which leads to the lower values of stabilized entanglement, is even more obvious 
because in these examples we consider the relatively large values of $Q$. Due to the enhanced two-mode squeezing which also intensifies the effects of the noise terms, the existence of any type of noise can stabilize the entanglement in contrast to the constant growth of entanglement by assuming an evolution without the influence from the noises. The balance of the intensified decoherence and optomechanical coupling is thus seen to stabilize the entanglement in the dynamically unstable regime, where a blue-detuned CW drive enhances the squeezing action on the cavity and mechanical modes. 

The larger oscillation amplitude for a realized entanglement under the stronger drive, as in Figs. 2 and 4, 
exists to the entanglement stabilized under any type of the noises. From the examples in Fig. 6, we find that a stronger noise effect actually displaces the entanglement values under fix drive intensity $E$ to the lower side along the vertical axis. With the combined effect from both the cavity and mechanical noises, the realized optomechanical entanglement, as the plots moved towards the horizontal axis, can exhibit periodic ESD and entanglement revival; see the black curves in Figs. 6(a) and 6(c). For clearer illustrations of the balanced decoherence and optomechanical coupling, which affect the evolution of our concerned entanglement, we apply the further increased drive power to one of the setups considered in Fig. 6. 
The results of their evolutions are given in Fig. 7. Except for the obvious change at the beginning steps of raising the drive intensity (the two figures on the first row of Fig. 7), the evolution of the entanglement varies little in spite of adding more drive power successively. The effects of creating and killing the entanglement, both of which become more significant with increased drive power, can well balance each other over a considerable range of the drive intensity.

\section{Condition for realizing high-temperature entanglement}
\label{sec:rmts}

The most interesting issue for our concerned optomechanical entanglement is how it could survive at high temperature. 
For a simple quantum mechanical oscillator under thermal decoherence, which has the number of coherent oscillations as $\omega_m/(\gamma_m n_{th})=Q f_m \frac{h}{k_B T}$ ($f_m=\frac{\omega_m}{2\pi}$), its decoupling from the thermal environment's influence should satisfy the condition $Q/n_{th}\gg 1$, where $n_{th}=\frac{k_BT}{\hbar\omega_m}$ for the oscillator at high temperature. Under this condition the purity of the quantum states of an OMS may be preserved, as it was mentioned in the previous studies (see, e.g. \cite{boundary3, deen1}). To the quantum entanglement of the OMS, however, an extra factor is the coupling between the cavity and mechanical modes as the mutual interaction to create the entanglement. As we have understood from the above discussions, the stabilized entanglement due to blue-detuned CW drive is the result of the balanced decoherence and mutual interaction, both of which become larger with time in the concerned regime of dynamical instability. Could such a balance be reset by a proper optomechanical coupling so that the entanglement can appear with the lower ratio $Q/n_{th}$ for the systems?

\begin{figure}[b!]
\centering
\includegraphics[width=\linewidth]{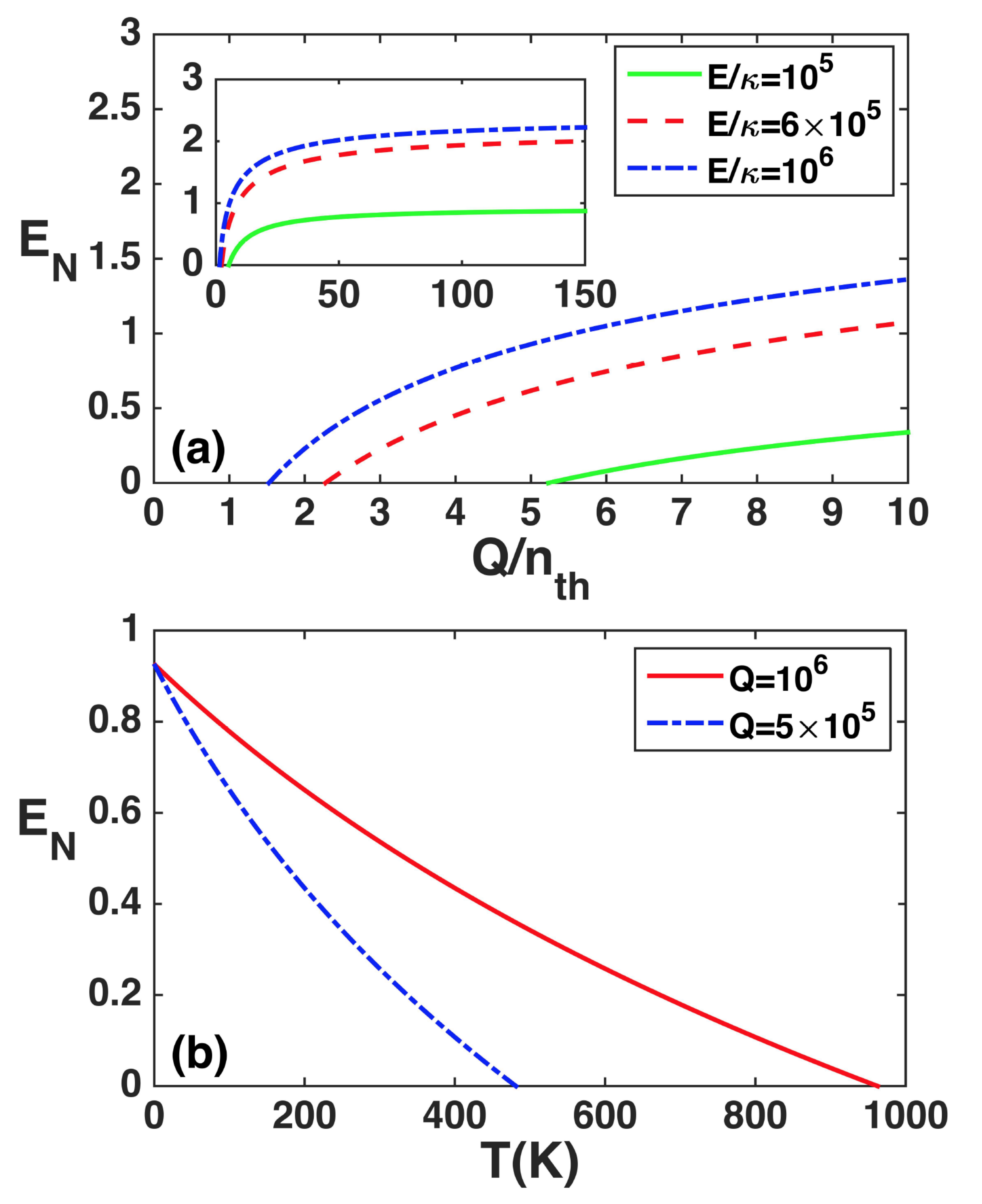}
\caption{Relations between the achieved peak value entanglement with the temperature and mechanical quality factor. 
(a) The ratio $Q/n_{th}$ for the existence of optomechanical entanglement. For all the given drive intensities, the entanglement appears with the ratio in the order of $Q/n_{th}\sim 1$. 
The inset shows a saturation of the values of the stabilized entanglement. With the even higher drive intensities, 
the entanglement may exist at a point $Q/n_{th}<1$. (b) The tendency of the stabilized peak values of entanglement, which are obtained given the drive intensity $E/\kappa=10^5$, with the temperature up to the degree higher than $900$ K for a technically available quality factor $Q=10^6$. 
In both (a) and (b) we have $\omega_m/2\pi=100$ MHz, and $g_m/\kappa=10^{-4},\omega_m/\kappa=10$, $\Delta/\omega_m=-1$ as in Fig. 3.}
\end{figure}

This conjecture can be solved by finding the direct relation of the evolved entanglement with the ratio $Q/n_{th}$. Considering the possible appearance of ESD and entanglement revival as in Fig. 4 and Figs. 6(a) and 6(c), we here use the stable peak value of the entanglement to indicate its existence. The setup in Fig. 3 is used to illustrate the entanglement in general thermal environment, with a flexibility that its mechanical quality factor can be adjusted in preparing the setup.
The results in Fig. 8(a) show that the optomechanical entanglement can appear with a proper drive intensity as long as the concerned ratio is in the order $Q/n_{th}\sim 1$, which is much more relaxed than the condition $Q/n_{th}\gg 1$ for an OMS to be decoupled from thermal decoherence. The modified balance between decoherence and optomechanical coupling due to the change of system parameters in the evolution of entanglement can be seen in Fig. 8(a). For instance, at the fixed point $Q/n_{th}=2$, no entanglement exists under the drives of lower intensity ($E/\kappa=10^5, 6\times 10^5$), but a higher intensity ($E/\kappa=10^6$) inducing a more significant coupling than the simultaneously intensified decoherence can give rise to a non-zero entanglement up to the order of $E_\mathcal{N}=0.1$. As seen from the inset of Fig. 8(a), a very high ratio $Q/n_{th}$ cannot improve on the entanglement forever because of the decoherence from the cavity reservoir which takes an independent action on the systems. In terms of the temperature of the thermal reservoir, one can illustrate the general tendency of the entanglement as in Fig. 8(b). It is possible to have entanglement at much higher than room temperature, 
given the technically achievable quality factor $Q$ as those in Fig. 8(b). 
Unlike the previous proposals (see, e.g. \cite{cla1,cla2,cla3,cla4,cla5, cla7,cla8,cla9}), 
such robust entanglement should be realized with dynamically unstable systems. 

A further question is how to estimate the necessary interaction (optomechanical coupling) for the realization of the entanglement at a certain temperature. The reciprocal of the 
ratio $Q/n_{th}$ is proportional to the factor $(\gamma_m/\kappa) n_{th}$, which indicates the thermal decoherence rate from the mechanical reservoir. The realized entanglement is the result of the system dynamical evolution determined by the effective optomechanical couplings as well as by the decoherence from the reservoirs. An intuitive model for understanding the main features of the system dynamics is the one obtained in the limit of $\omega_m/\kappa\rightarrow \infty$, whose dynamical equations are given as Eq. (\ref{eq:sdm}). After rewriting the dynamical equations in terms of the dimensionless time scale $\kappa t$, one will see that all evolved quantities according to the equations change with only three parameters $J/\kappa$, $\gamma_m/\kappa$ and $n_{th}$. Under the condition $\gamma_m/\kappa\ll 1$ that can be achieved for the concerned OMSs, the finally stable entanglement in this limit can also be given as an analytical form in 
Eq. (\ref{analy}).
Then we can clearly plot how the realized stable entanglement distributes in a two-dimensional parameter space as in 
Fig. 9. There is a clear-cut boundary $\gamma_m n_{th}=g_mE/\omega_m$ for the existence of entanglement, as the diagonal line in this figure. For a fixed thermal decoherence at the rate $(\gamma_m/\kappa) n_{th}$, 
the entanglement can be obtained with a certain drive power that realizes the proper coupling intensity, reducing the existence of optomechanical entanglement to a simple condition 
\begin{align}
\gamma_m n_{th}<J=g_mE/\omega_m. 
\label{relation}
\end{align}
Due to the possible adjustment of the effective coupling intensity $J$, optomechanical entanglement satisfying the condition can be realized even with a ratio $Q/n_{th}<1$. The OMSs with $\omega_m/\kappa\gg 1$, which are relevant to the relation, have been experimentally realized thus far (see, e.g. \cite{cool7}).     

In the realistic situations of finite sideband resolution $\omega_m/\kappa$, the boundary in Fig. 9 will be deformed for the small parameters on the left lower side. 
Though a simple mathematical expression for the realistic boundary does not exist, one can numerically determine the specific relations between the entanglement and the given system parameters. 
As an example, the system with $g_m/\kappa=10^{-4}$ and driven by a blue-detuned CW drive with $\Delta/\omega_m=-1$ and $E/\kappa=10^7$ can have the averaged stable entanglement of $E_\mathcal{N}\approx 0.093$, when the mechanical resonator having $\omega_m/\kappa=10$ and $\gamma_m/\kappa=1/900$ is in a thermal environment corresponding to $n_{th}=10^4$. 
The ratio $Q/n_{th}$ in this example is only $0.9$.

\vspace{0cm}
\begin{figure}[t!]
\centering
\includegraphics[width=0.9\linewidth]{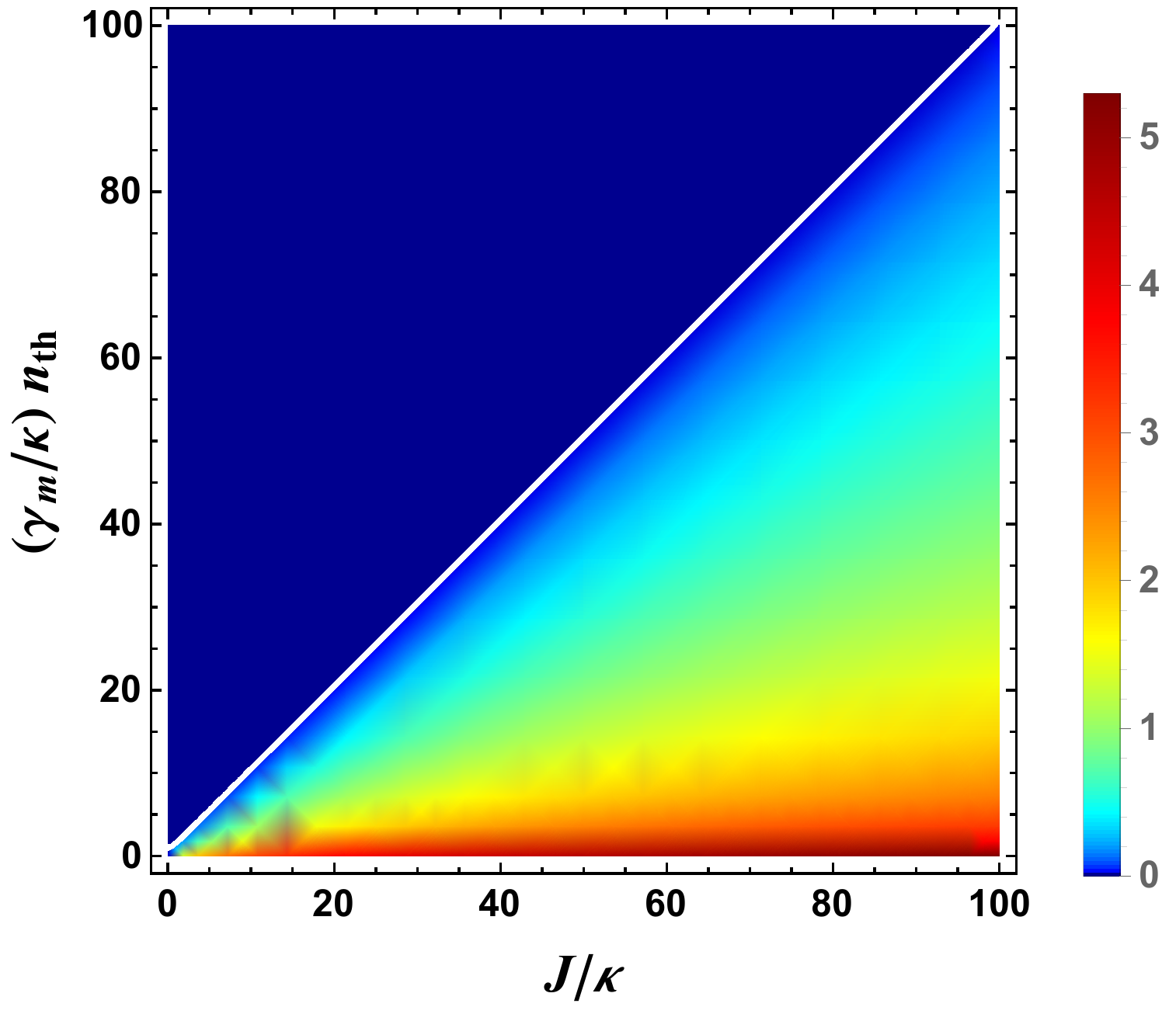}
\caption{Stabilized value of entanglement in the space of the parameters $J/\kappa=g_mE/(\omega_m \kappa)$ and $(\gamma_m/\kappa)n_{th}$. The result is obtained with Eq. (\ref{analy}) which comes from the system dynamics simplified 
with $\omega_m/\kappa \gg1$ and $\gamma_m/\kappa\ll 1$. The optomechanical coupling intensity on the general level is indicated by the first parameter, and the general thermal decoherence rate is measured by the second parameter.}
\end{figure}
\vspace{-0.5cm}

\section{Discussion and Conclusion}
\renewcommand{\thefigure}{S-\arabic{figure}}
\renewcommand{\theequation}{S-\arabic{equation}}
\setcounter{equation}{0}
\setcounter{figure}{0}
\label{sec:conclusion}

Quantum properties at room temperature are important to both fundamental researches and potential applications. Recently, the existence of such quantum properties has been experimentally investigated with OMSs \cite{t-1,t-2}. In these experiments the quantum fluctuations are interpreted as those around the steady states of a system. In the current work we have presented a general method to describe the associated fluctuations around arbitrary motions of weakly coupled OMSs (their time-dependent 
cavity mean field and mechanical average displacement), which can find applications to different regimes that are relevant to experimental research, especially those without dynamical stability. Entanglement of OMSs is a widely interested phenomenon concerning the macroscopic quantum states of these systems. Previously, due to the lack of a suitable method to describe the entanglement in dynamically unstable regimes, the proposals for making the observable optomechanical entanglement were mostly confined to red-detuned CW drives or blue-detuned pulses. In an experimental realization \cite{entangle}, ultra-low temperature is required, and an additional procedure of cooling the mechanical resonator should be applied before entangling it with another blue-detuned pulse field. However, blue-detuned CW drive field can give rise to much more 
significant two-mode squeezing effect for creating the concerned entanglement. Our analysis shows that the entanglement due to blue-detuned CW drive can become stable in thermal environment, indicating the possibility of realizing such entanglement even at very high temperature. The condition for obtaining the quantum optomechanical entanglement is found to be much more relaxed than the well-known criterion for preserving an OMS's coherence (such as the purity) in a general thermal environment. Direct applying blue-detuned CW drives will also significantly simplify the generation of such optomechanical entanglement. All these conclusions are based on the single prerequisite that the quantum state of a mechanical resonator at any temperature of the environment can be initially prepared as the one in thermal equilibrium, i.e. the initial state given by Eq. (\ref{ini}).

A technical concern in the proposed dynamical unstable regime 
is the continually increasing displacement of the mechanical resonator. After the displacement proportional to the real part of $\langle\hat{b}(t)\rangle$ is beyond an extent so that the effective linear dynamics begins to be invalid, the mechanical motion would turn into a self-induced oscillation due to the saturation of nonlinearity \cite{opc5}. The entanglement between the cavity field and mechanical resonator can still exist after reaching the phase, though the effective dynamics should be modified to incorporate the associated nonlinear effects and the quantification of the entanglement with the logarithmic negativity will become impossible due to the deviation from Gaussian states. Proper material could be chosen for the mechanical resonator to realize the sufficient time for our concerned linear dynamical evolutions. In the current study we also work with the white noise model for the mechanical reservoir as in \cite{opc5}. The more realistic colored noise of thermal reservoir can be considered in the effective dynamical equations Eq. (\ref{eq:dm}) to correct the calculated CM elements. With a narrow spectrum of the mechanical noise taking the actual effect, the white noise model should be sufficient for studying the dynamical evolutions \cite{book}.

The fundamental meaning of the current study is to understand how quantum entanglement can exist in 
thermal environment through the examples of OMSs at different temperatures. These examples demonstrate that quantum systems' coherence and entanglement are totally different. The loss of the latter due to the environmental noises can be offset by the mutual interaction within the systems, so that there exist richer phenomena beyond the previously envisioned general scenario of ESD. Together with the numerical simulations in the general situations, a quantitative relation Eq. (\ref{relation}) obtained in the highly resolved sideband regime for OMS indicates how large the interaction should be required for preserving the entanglement at any temperature. Similar results are expected to find in other physical systems.

\begin{figure}[h!]
\centering
\includegraphics[width=\linewidth]{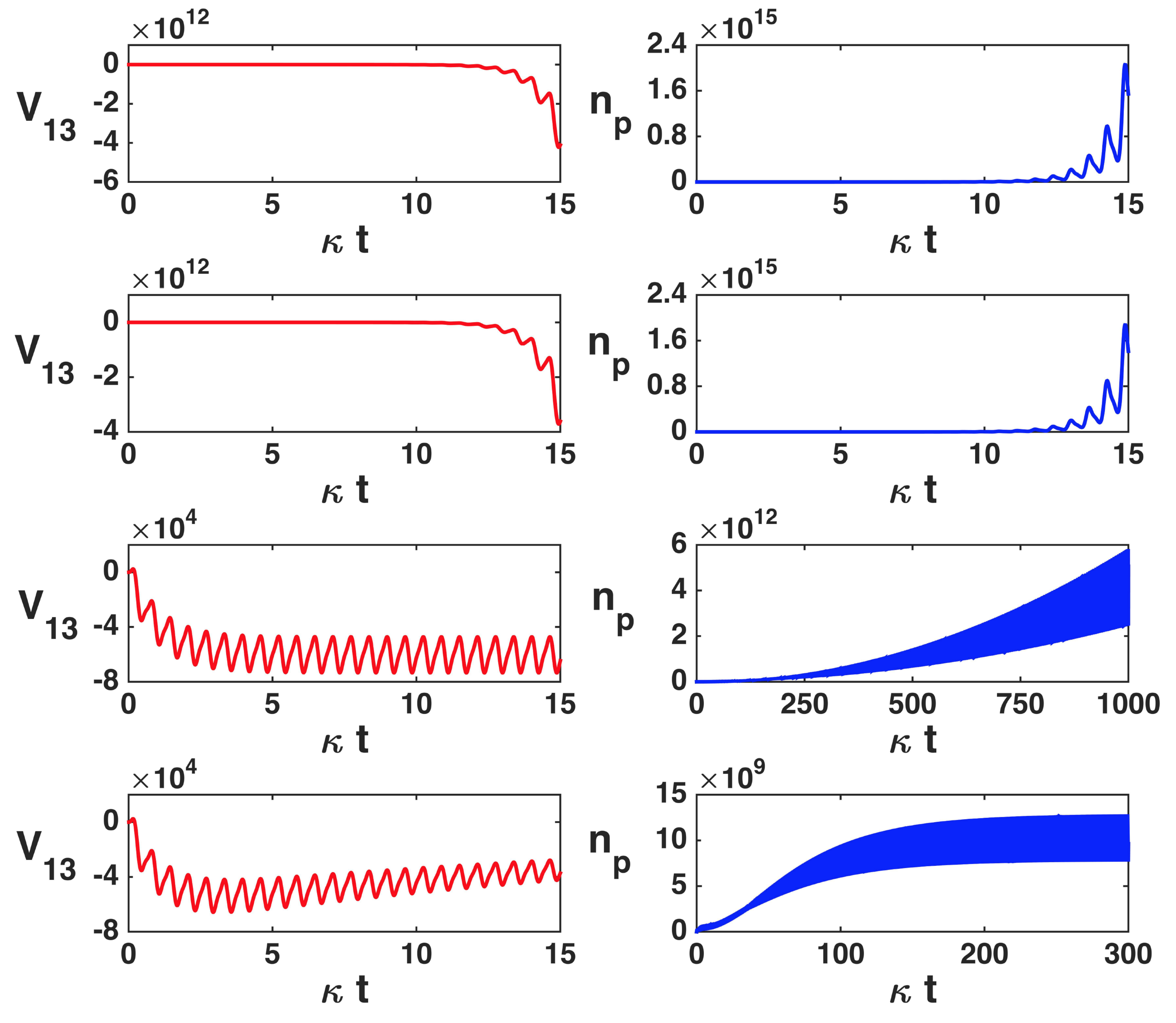}
\caption{Evolutions of a CM matrix element and the associated cavity photon number. The intensity of the first drive is fixed at $E_1=10^5\kappa$, and that of the second drive is taken as $E_2=0$, $10^4\kappa$, $10^5\kappa$ and $1.01\times 10^5\kappa$ from the upper to the lower rows. The system parameters are given as $g_m=10^{-4}\kappa$, $\omega_m=10\kappa$, $\gamma_m=10^{-4}\kappa$, $-\Delta_1=\Delta_2=10\kappa$, and $n_{th}=4\times 10^4$.}
\end{figure}
\vspace{-0cm}

\section*{Appendix: a scheme for verifying the associated dynamics}
\label{sec:detection}

A related issue is how to know that the entanglement of OMSs being driven by blue-detuned CW field evolves according to 
the illustrated patterns. In the previous studies, the degree of the entanglement of two light fields is inferred from the measured elements of the CM defined in Eq. (\ref{CM}) \cite{detect1, detect2}. Such CM elements involving the mechanical mode, however, cannot be directly measured. In one experimental realization of the OMS entanglement with a pulsed field \cite{entangle}, another red-detuned pulse is applied to swap the optomechanical entanglement to that between two light fields, 
whose correlations can be detected. In our concerned dynamically unstable regime due to blue-detuned CW drive, the magnitudes of the CM elements keep growing with time. Though it is possible to implement a real-time homodyne detection of light fields 
\cite{detect3}, a direct determination of the CM elements in the currently concerned regime of dynamical instability is rather challenging.  

Here we propose a scheme to verify if the system will evolve according to the illustrated effective dynamics so that the predicted optomechanical entanglement can exist accordingly. It is to add one more CW drive field of red-detuned at $\Delta_2=\omega_m$, which acts simultaneously with the blue-detuned one ($\Delta_1=-\omega_m$)
for realizing the entanglement with the mechanical resonator. 
The two drive fields can have different polarizations, so that their respectively induced cavity fields can be discriminated in detection. After a linearization procedure similar to that in Sec. II of the main text, the effective dynamical equations become
\begin{align}
\frac{d}{dt}\hat{\vec{d}}(t)=\hat{M}(t)\hat{\vec{d}}(t)+\vec{\lambda}(t)+\hat{\vec{n}}(t),
\label{dynamic-3mode}
\end{align}
where
the system mode vector $\hat{\vec{d}}=\big(\hat{a}_1,\hat{a}_1^\dagger,\hat{b},\hat{b}^\dagger,\hat{a}_2,\hat{a}_2^\dagger\big)^T$ has included the two cavity modes $\hat{a}_1$ and $\hat{a}_2$ from the different drives. The coherent drive terms are 
in the vector $$\vec{\lambda}=\big(\lambda_1, \lambda_1^\ast, \lambda_2, \lambda_2^\ast, \lambda_3,\lambda_3^\ast\big)^T$$ with $\lambda_{1(3)}(t)=-\kappa E_{1(2)}(t)$ and 
$\lambda_2(t)=i g_m e^{i\omega_m t}|E_1(t)|^2+i g_m e^{i\omega_m t}|E_2(t)|^2$, with $E_i(t)=\frac{iE_i}{\Delta_i}(1-e^{i\Delta_i t})$ for $i=1,2$.
The noise drive terms are correspondingly given in
$$\hat{\vec{n}}=\big(\hat{n}_1,\hat{n}_1^\dagger,\hat{n}_2,\hat{n}_2^\dagger,\hat{n}_3,\hat{n}_3^\dagger\big)^T$$ with 
$\hat{n}_{1(3)}(t)=\sqrt{2\kappa}\hat{\xi}_c (t)$ and $\hat{n}_2(t)= \sqrt{2\gamma_m}\hat{\xi}_m(t).$ 
The dynamics matrix in the above equations reads
\begin{widetext} 
\begin{align}
\hat{M}(t)=\begin{pmatrix}
 -\kappa& 0 & ig_mE_1(t) e^{-i\omega_m t} & ig_mE_1(t) e^{i\omega_m t} & 0 & 0 \\
 0& -\kappa & -ig_mE_1^*(t) e^{-i\omega_m t} & -ig_mE_1^*(t) e^{i\omega_m t} & 0 & 0\\
ig_mE_1^*(t) e^{i\omega_m t} &ig_mE_1(t) e^{i\omega_m t}  & -\gamma_m & 0 &ig_mE_2^*(t) e^{i\omega_m t} &ig_mE_2(t) e^{i\omega_m t}\\
 -ig_mE_1^*(t) e^{-i\omega_m t} & -ig_mE_1(t) e^{-i\omega_m t}  & 0  & -\gamma_m &  -ig_mE_2^*(t) e^{-i\omega_m t} & -ig_mE_2(t) e^{-i\omega_m t} \\
 0 & 0 & ig_mE_2(t) e^{-i\omega_m t} & ig_mE_2(t) e^{i\omega_m t} & -\kappa & 0\\
 0 & 0 & -ig_mE^*_2(t) e^{-i\omega_m t} & -ig_mE^*_2(t) e^{i\omega_m t} & 0 & -\kappa
\end{pmatrix}.
\end{align}
\end{widetext}

\vspace{0cm}
\begin{figure}[t!]
\centering
\includegraphics[width=\linewidth]{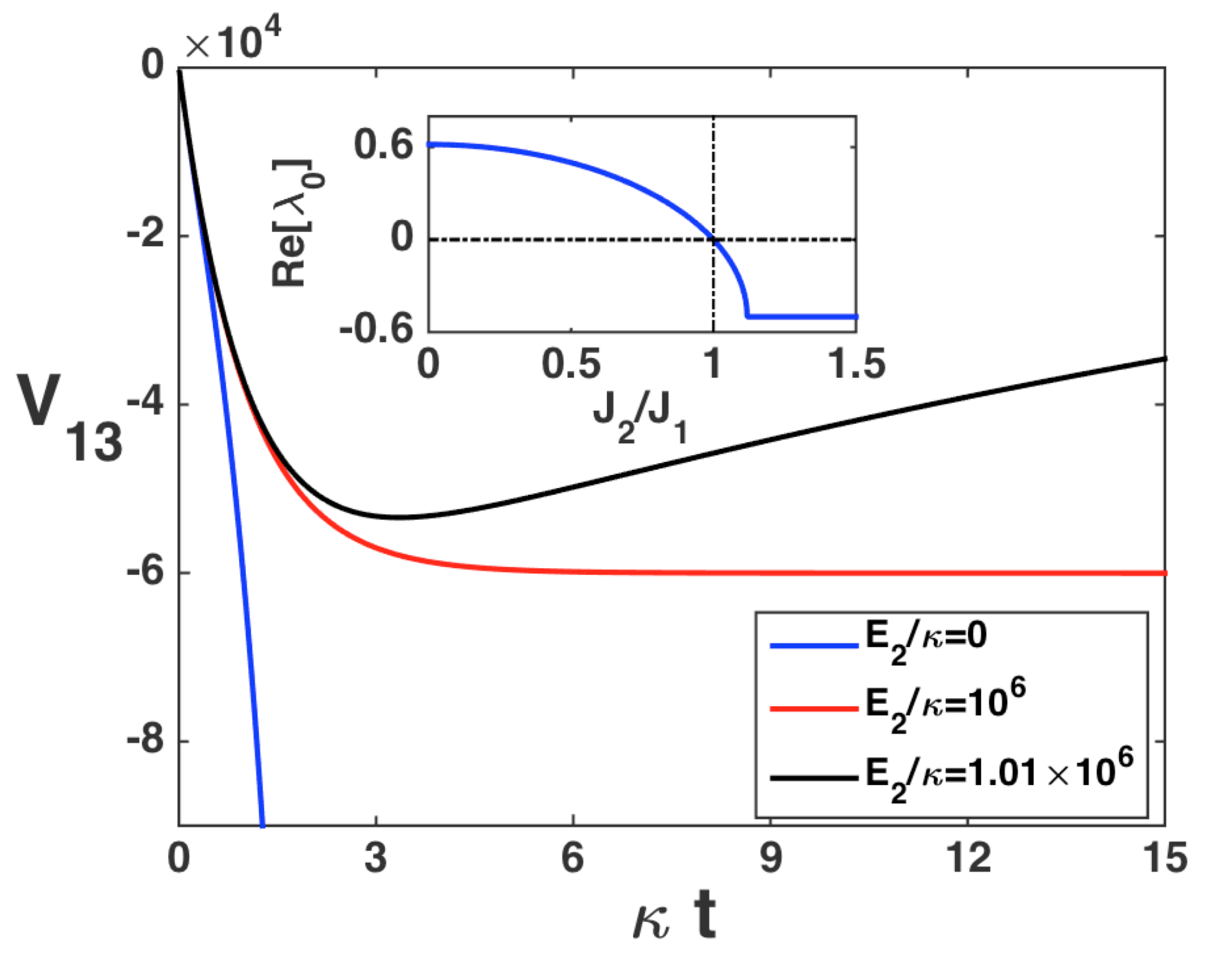}
\caption{Evolutions of a CM element under the different intensities of the 
second drive field. Here we illustrate the evolution patterns for a setup of 
highly resolved sideband $\omega_m/\kappa\rightarrow \infty$ as that described 
by Eq. (\ref{eq:sdm}). Here $J_i=g_mE_i/\omega_m$ for $i=1,2$. The inset shows the tendency of the real part of an eigenvalue of the dynamics matrix $\hat{M}$, which turns from being positive to being negative, indicating that the system can undergo the transition from dynamical instability to dynamical stability.}
\end{figure}
\vspace{-0cm}

The corresponding Gaussian states of the system, which evolve according to the above dynamical equations, can be fully described by a correlation matrix $\hat{V}$, with its elements defined as 
\begin{eqnarray}
V_{ij}(t)=1/2\langle \delta \hat{u}_i(t) \delta \hat{u}_j(t)+\delta \hat{u}_j(t)\delta \hat{u}_i(t)\rangle,
\label{elements}
\end{eqnarray} 
where $$\delta \hat{u}_i(t)=\hat{u}_i(t)-\langle \hat{u}_i(t)\rangle$$ 
for $\hat{\vec{u}}(t)=(\hat{x}^1_c(t),\hat{p}^1_c(t),\hat{x}_m(t),\hat{p}_m(t),\hat{x}^2_c(t),\hat{p}^2_c(t))^T$.
The fluctuations $\delta \hat{u}_i(t)$ can be around arbitrary motion $\langle \hat{u}_i(t)\rangle$ of the system, being more general than those around stable oscillations as considered previously \cite{w-c}. The system dynamics with $E_2=0$ (Eq. (\ref{eq:dm}) in the main text) is mapped to the another one described by Eq. (\ref{dynamic-3mode}) after adding one more light field. The different linear dynamical evolutions with $E_2=0$ or $E_2\neq 0$ have one-to-one correspondence. 
Our purpose is to infer the properties of the system described by Eq. (\ref{eq:dm}) with those observed from the situations of $E_2\neq 0$.    

The introduction of the second drive will make the evolution of cavity photon number $n_p$ become different, and the evolved photon numbers are also in one-to-one correspondence with the evolved CM elements, as seen from the comparisons in Fig. S-1. There, with the increased intensity of the second drive field, both the CM elements (with the example of $V_{13}$ between an optical mode and the mechanical mode) and the cavity photon number will change from unstably growing to stably oscillating. 
Such change can be seen more clearly from Fig. S-2, which depicts the evolutions of 
a CM element for the simplified system with $\omega_m/\kappa\rightarrow \infty$. 
The change of photon number evolution patterns due to different intensities of the second red-detuned drive field can be found with detection. Then one can deduce how the elements of CM of the system will evolve 
from the one-to-one correspondence between the photon number and CM elements such as that in Fig. S-1---if the photon number is found to become stable exactly under increased drive intensity $E_2$, the corresponding evolutions of the CM elements, including those in the dynamically unstable regimes, can be indirectly confirmed. Once the correlations of the time-dependent fluctuations defined after Eq. (\ref{elements}) can be directly measured with more advanced technology, 
the existence and magnitude of the concerned entanglement will be better demonstrated. 

\vspace{0.5cm}
\begin{acknowledgments}
This work is funded by National Natural Science Foundation of China (Grant No. 11574093 and No. 61435007); 
Natural Science Foundation of Fujian Province of China (Grant No. 2017J01004); Promotion Program for Young and Middle-aged Teacher in Science and Technology Research of Huaqiao University (Grant No. ZQN-PY113). Q. L. acknowledges the support by the Training Program of Fujian Excellent Talents in University. This research is also supported by the Arkansas High Performance Computing Center and the Arkansas Economic Development Commission. 
\end{acknowledgments}

\end{document}